\documentclass[journal]{IEEEtran}

\usepackage[left=1.5cm, right=1.5cm, top=1.5cm, bottom=1.5cm]{geometry}
\usepackage{amsmath}
\usepackage{graphicx} 
\usepackage{dblfloatfix}  
\usepackage{multicol}
\usepackage{array}
\usepackage{mathrsfs}
\usepackage{multirow}
\usepackage{xcolor}
\usepackage{comment}
\usepackage[export]{adjustbox}
\usepackage{subcaption}
\usepackage{enumitem}
\usepackage{verbatim}
\usepackage{algorithm,algpseudocode}

\algnewcommand{\IfThenElse}[3]{
  \State \algorithmicif\ #1\ \algorithmicthen\ #2\ \algorithmicelse\ #3}
  
\usepackage{lipsum}
\usepackage{titlesec}

\newcommand{\alphalphval}[1]{.\arabic{equation}}
\newcolumntype{P}[1]{>{\centering\arraybackslash}p{#1}}

\AtBeginDocument{%
  \AtBeginEnvironment{subequations}{%
  }
}

\titlespacing\section{0pt}{4pt plus 0pt minus 2pt}{4pt plus 0pt minus 2pt}
\titlespacing\subsection{0pt}{4pt plus 0pt minus 2pt}{4pt plus 0pt minus 2pt}
\titlespacing\subsubsection{0pt}{4pt plus 0pt minus 2pt}{4pt plus 0pt minus 2pt}

\usepackage[font=small]{caption}

\usepackage{amssymb}

\usepackage{accents}

\usepackage{booktabs}
\usepackage{comment}
\usepackage[utf8]{inputenc}
\usepackage{nomencl}
\makenomenclature
\usepackage{etoolbox}
\renewcommand\nomgroup[1]{%
    \item[\bfseries
    \ifstrequal{#1}{A}{Indices and Sets}{%
    \ifstrequal{#1}{B}{Indices}{%
    \ifstrequal{#1}{C}{Parameters}{%
    \ifstrequal{#1}{D}{Variables}{%
    \ifstrequal{#1}{E}{Abbreviations}{}}}}}%
    ]}
\makeindex

\usepackage{amsmath}
\usepackage{amssymb}

\usepackage{nomencl}
\usepackage{utfsym}
\usepackage{algpseudocode}
\algrenewcommand\algorithmicrequire{\textbf{Input:}}
\algrenewcommand\algorithmicensure{\textbf{Output:}}
\makenomenclature
\usepackage{nomencl}
\usepackage[english]{babel}

\usepackage{amsthm}
\usepackage{xcolor}
\theoremstyle{definition}

\theoremstyle{theorem}
\newtheorem{theorem}{Theorem}
\makenomenclature

\begin{document}
\renewcommand{\qedsymbol}{}
\title{A Conformal Prediction-Based
Chance-Constrained Programming Approach for 24/7 Carbon-Free Data Center Operation Scheduling }


\author{Yijie Yang,~\IEEEmembership{Graduate 
 Student Member,~IEEE,}
	Jian Shi,~\IEEEmembership{Senior Member,~IEEE,}
		  Dan Wang,~\IEEEmembership{Senior Member,~IEEE,}
		Chenye Wu,~\IEEEmembership{Senior Member,~IEEE,}
		and Zhu Han,~\IEEEmembership{Fellow,~IEEE}
		\vspace{-7mm}
	}
	
\maketitle

	
     \begin{abstract}
The rapid growth of AI applications is dramatically increasing data center energy demand, exacerbating carbon emissions, and necessitating a shift towards 24/7 carbon-free energy (CFE).  Unlike traditional annual energy matching, 24/7 CFE requires matching real-time electricity consumption with clean energy generation every hour, presenting significant challenges due to the inherent variability and forecasting errors of renewable energy sources.  Traditional robust and data-driven optimization methods often fail to leverage the features of the prediction model (also known as contextual or covariate information)  when constructing the uncertainty set, leading to overly conservative operational decisions. 
This paper proposes a comprehensive approach for 24/7 CFE data center operation scheduling, focusing on robust decision-making under renewable generation uncertainty.  This framework leverages covariate information through a multi-variable conformal prediction (CP) technique to construct statistically valid and adaptive uncertainty
sets for renewable forecasts. The uncertainty sets directly inform the chance-constrained programming (CCP) problem, ensuring that chance constraints are met with a specified probability. We further establish theoretical underpinnings
connecting the CP-generated uncertainty sets to the statistical feasibility guarantees of the CCP. Numerical results highlight the benefits of this covariate-aware approach, demonstrating up to
6.65\% cost reduction and 6.96\% decrease in carbon-based energy usage compared to conventional covariate-independent methods, thereby enabling data centers to progress toward 24/7 CEF.

	\end{abstract}
	
	\begin{IEEEkeywords}
		Carbon Emission; Data Center Energy System; Robust Optimization; Chance-constrained Optimization; 24/7 Carbon-Free.
  
	\end{IEEEkeywords}

\section{Introduction}
\subsection{Motivation and Challenges}
The surge in AI applications since the release of ChatGPT in 2022, with computational power doubling approximately every 100 days and AI tasks consuming significantly more energy than traditional computation tasks \cite{carbonDC}, is dramatically increasing electricity demand. Global data center electricity demand is projected to increase by 50\% by 2027 and potentially 165\% by the decade's end \cite{AIdemand}. Consequently, carbon emissions from tech giants like Microsoft, Meta, and Google have substantially increased, largely due to AI investments \cite{AIcomcarbon}.

To mitigate the environmental impacts of AI-driven energy demand, increasing attention is being directed toward achieving 24/7 Carbon-Free Energy (CFE). Unlike traditional renewable energy procurement, where companies purchase unbundled energy attribute certificates (EACs) to match annual consumption volumes, 24/7 CFE requires matching electricity demand with carbon-free energy generation \textit{every hour}, in every region where operations occur \cite{247UN}. It ensures that the clean energy supply aligns with real-time consumption, reducing reliance on carbon-intensive grid electricity during periods when renewables are unavailable. Major technology companies have embraced this shift: Google has committed to operating entirely on 24/7 carbon-free energy globally by 2030~\cite{googleCEF}, while Microsoft aims to become carbon negative by 2030 through a combination of emissions reductions and carbon removal~\cite{MicroCEF}. Meanwhile, Amazon, Meta, and Google have collectively invested in 22 GW of renewable energy capacity to support their Net Zero goals \cite{acun2023carbon} by relying on locally generated renewable energy in complement to the electricity provided by the local grid.

However, achieving true 24/7 CFE introduces two unique challenges. First, the variability and intermittency of renewable energy sources, such as solar and wind, make it difficult to guarantee continuous clean power availability. This variability cannot be addressed solely by scaling up renewable infrastructure; instead, it necessitates a shift in how data centers manage their operations. Fortunately, a significant fraction of data center workloads, such as AI training tasks, are latency-tolerant and governed by service level objectives (SLOs) that allow flexible execution timelines. For example, Google reports 30-40\% with a 24-hour SLO, Meta indicates 20-30\% with varying SLOs, and Microsoft Azure identifies 70\% as delay-tolerant \cite{xing2023carbon}. This latent flexibility offers a critical opportunity: by implementing emission-aware scheduling, data centers can dynamically align their power consumption with periods of renewable energy abundance, with the goal of closing the gap between supply and demand and advancing toward 24/7 CFE \cite{acun2023carbon}. 
\vspace{-0.5mm}

The second major challenge in realizing 24/7 CFE lies in managing the uncertainty inherent in forecasting future renewable energy availability.   Although forecasting models,  typically based on historical data and machine learning techniques, can provide estimates \cite{ruan2020review}, they inevitably introduce errors, especially under rapidly changing environmental conditions \cite{zhang2018robustly}.  These prediction errors pose a threat to the reliability of scheduling strategies, as misalignment between expected and actual clean energy availability can result in missed decarbonization opportunities or even system instability. Traditional robust optimization methods address uncertainty but often rely on static or heuristically sized uncertainty sets that do not leverage real-time contextual information (covariates such as weather, weekday, month, and
season information, etc.) \cite{yang2021robust, ciftci2019data}. This oversight leads to overly conservative decisions in the downstream optimization process. The core challenge is developing statistically rigorous and adaptive uncertainty modeling approaches that can dynamically adjust based on real-time covariates, providing a more accurate and less conservative representation of future renewable availability for the downstream operation problem.


In the face of these two challenges, this paper focuses on addressing two critical research questions: 1) \textit{How to model the energy consumption of a data center so we can match the workload with low-carbon periods to achieve 24/7 CEF?} 2) \textit{How to make robust operation and further guarantee the 24/7 renewable energy utilization utilizing the covariate information?}

\subsection{Literature Review}
\textit{1) 24/7 Renewable Energy Operation of Data Centers.} Several approaches have been proposed to enhance the environmental and economic performance of green data centers. Ghamkhari \textit{et al.} focuses on maximizing the profit of green data centers by incorporating and managing onsite renewable energy generators in \cite{ghamkhari2013energy}.
Chen \textit{et al.} directly addresses carbon emissions by optimizing the energy management strategies for distributed data centers operating under demand response (DR) programs in \cite{chen2013electric}.
Kwon \textit{et al.} investigate the co-optimization of server operations and power procurement in \cite{kwon2018demand}. Their approach utilizes renewable energy sources and energy storage systems to facilitate power procurement and enable demand response capabilities.
Kiani \textit{et al.} introduce a model that differentiates workloads into green and brown categories in \cite{kiani2016profit}. This decomposition allows for considering the varying costs and environmental impacts associated with green and brown energy, enabling the strategic allocation of green workloads to data centers based on the localized availability and cost of green energy, with the overarching goal of profit maximization. The solution for achieving 24/7 carbon-free data centers operations is first introduced in \cite{acun2023carbon}. It addresses investment strategies for 24/7 carbon-free operations using energy storage, renewable energy, and workload shifting, but focuses on a long-term horizon.

\textit{Research Gap:} While existing studies provide valuable insights into cost optimization, renewable integration, and investment strategies for low-carbon data centers, there is a lack of research addressing the operational-level implementation and modulation of 24/7 carbon-free energy solutions, which is critical for achieving ambitious future data centers sustainability goals.

\textit{2) Data-Driven Approaches for Constructing Uncertainty Sets.}  
Robust optimization (RO) is a well-established paradigm widely utilized to address inherent uncertainties, particularly those arising from intermittent renewable energy generation \cite{yang2021robust}. While commonly employed, traditional static uncertainty sets (e.g., box, budget, ellipsoidal) often yield overly conservative solutions by overemphasizing low-probability extreme scenarios. Furthermore, approaches that attempt to heuristically control uncertainty set size typically lack rigorous statistical guarantees. To mitigate this conservatism and introduce statistical rigor, research has explored integrating RO with chance-constrained programming (CCP). Specifically, data-driven CCP methods aim to construct uncertainty sets that ensure constraint satisfaction with a pre-defined probability under the worst-case distribution \cite{ciftci2019data, zhai2022data}, thus offering a balance between robustness and optimality.

\begin{figure}[t]
\centering
\includegraphics[width=0.8\columnwidth]{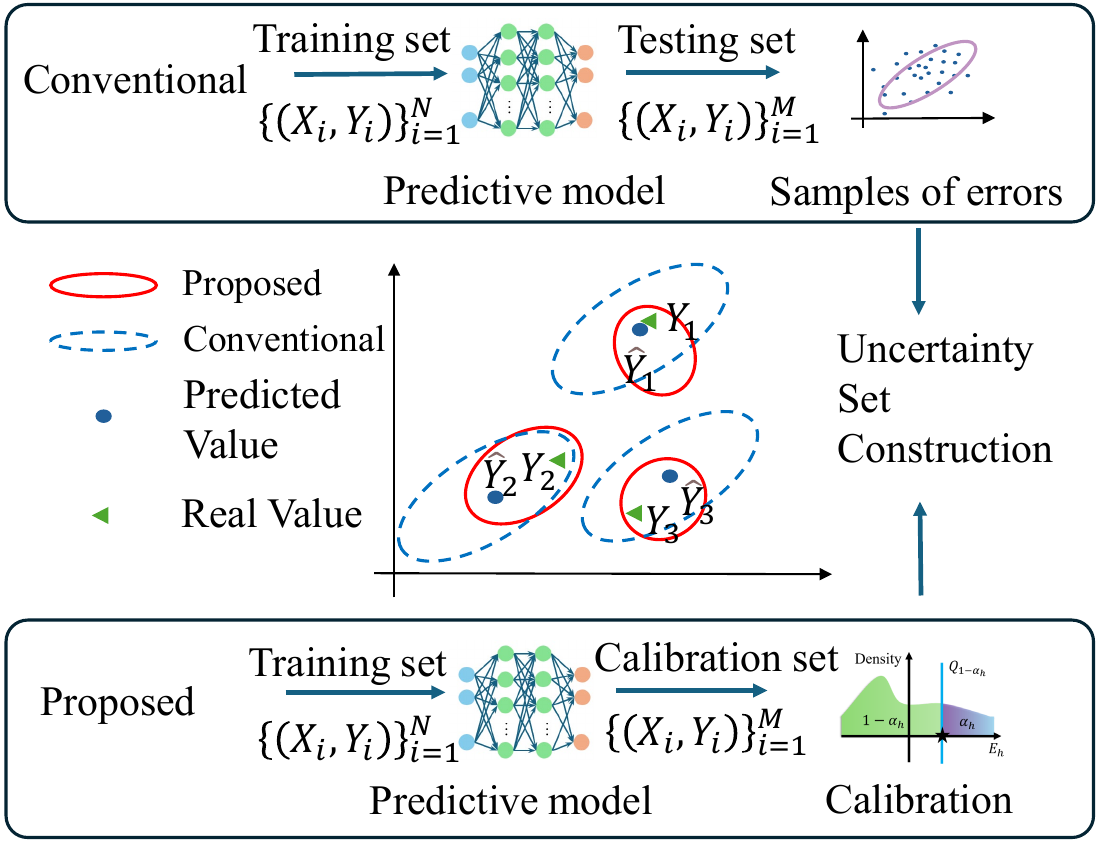}
\caption{Comparison between conventional and proposed  uncertainty set construction}
\label{fig:dif}

\end{figure}

Despite these advancements, a significant limitation of existing data-driven robust optimization approaches is that they are \textit{covariate-independent}~\cite{bertsimas2025constructing}. These methods typically use input features (i.e., covariates) such as wind speed, irradiance, temperature, or historical generation data to train a predictive model. However, the subsequent construction of the uncertainty set (e.g., intervals for RO or error distributions for CCP) is derived solely from the prediction errors observed in a test set, without dynamically leveraging the real-time covariate information at the prediction process. Therefore, as illustrated in Fig. \ref{fig:dif}, the size and shape of the generated uncertainty set often remain uniform across different new data points, despite varying underlying input conditions. This contrasts with an intuitive understanding that the level of uncertainty could change based on specific environmental or operational covariates. For the downstream optimization,  the conservative uncertainty set will increase total operational costs for data centers.

Some recent efforts have sought to incorporate covariate information into the dynamic sizing of uncertainty sets \cite{bertsimas2025constructing}. However, these methods are often restricted to specific prediction models (e.g., ensemble learning), involve computationally intensive dual theory or sample-based techniques, or are limited to addressing uncertainty solely in the objective function without providing feasibility guarantees for constraints \cite{sun2023predict}.

\textit{Research Gap}: Traditional robust and data-driven stochastic optimization methods overlook the direct relationship between exogenous input features (covariates) and the structure of prediction uncertainty. This oversight results in overly conservative decisions, as they fail to dynamically adapt the uncertainty set's characteristics based on real-time contextual information. Addressing this gap requires methods that explicitly integrate covariate information into the construction of statistically rigorous and adaptive uncertainty sets for optimization.

\subsection{Our Contributions}



In this study, we extend the application of conformal prediction by integrating it with chance-constrained programming. Specifically, quantile regression is employed to calibrate the uncertainty of the prediction model, generating finite-sample valid and conservative uncertainty sets with statistical feasibility guarantees. To further reduce the conservatism of classical pointwise conformal prediction, we propose a multi-variable conformal quantile prediction method, which improves the efficiency and practicality of uncertainty modeling. The numerical results demonstrate that the proposed methods achieve a cost reduction of up to 6.96\% and a reduction of 5.51\% carbon-based energy consumption compared to covariate-independent approaches, while maintaining statistical feasibility guarantees. The contributions of this study are summarized as follows:
\begin{enumerate}  
    \item We formulate a 24/7 CEF model designed specifically for optimizing data center operation scheduling, considering both flexible and inflexible load models while accounting for the dynamic carbon-based energy proportion of the power grid. 
    
    \item We propose a novel framework that integrates conformal prediction with chance-constrained programming to robustly optimize this model under renewable energy prediction uncertainty. This framework utilizes covariate information to enhance uncertainty modeling accuracy, thereby reducing operational cost and carbon-based energy of data centers.
    
    \item We establish theoretical guarantees that link the uncertainty sets generated by the upstream conformal prediction directly to the statistical feasibility of the downstream chance-constrained optimization problem, providing a theoretical connection between these two technologies.
\end{enumerate}

  The remainder of this paper is organized as follows: Section \ref{formulation} introduces the data center power consumption model and the workload scheduling model. Section \ref{robust_optimization} presents the chance-constrained problem, its reformulation, and the connection to conformal prediction. In Section \ref{conformal_prediction}, the conformal prediction approach for data center workload scheduling is detailed within the reformulated framework. Section \ref{Numerical_studies} provides the numerical studies, and Section \ref{conclusion} concludes the paper.

\section{Model Formulation} \label{formulation}
This section presents the modeling of a data center microgrid, followed by a description of the data center power consumption model, the carbon-based energy accounting model, the objective function, the energy storage system (ESS) model, and the power balance.  The energy procurement decision from the utility grid and the scheduling decision for onsite generation resources are co-optimized with the workload allocation decision. 

\subsection{Data Centers Power Consumption Model}
A data center typically comprises three primary components: IT equipment, refrigeration equipment, and the power distribution system \cite{yang2023two}. Power Usage Effectiveness (PUE) is a widely adopted metric for assessing data center energy consumption. PUE is defined as the ratio of the total power consumption of the data center to the power consumed solely by the IT equipment. Following the model in \cite{yang2023two} and \cite{zhou2022bi}, the total active power load of the data center can be determined as follows:
\begin{align}
Q_{t}^{\mathrm{DC} } & =\left[P^{\mathrm{idle}}+\left(PUE-1\right) P^{\text {peak }}\right] A_{ t}^{\mathrm{DC}} \notag\\
&+W_{t}^{\mathrm{DC}}\left(P^{\text {peak }}-P^{\mathrm{idle}}\right) / L^{\text {rate}}, \forall t  \in T \label{dc1}\\
& W_{t}^{DC} = \sum\nolimits_c^\mathcal{C} L_{c,t}^{FL}+L_{t}^{IFL}, \forall t \in T \label{dc2}\\
& A_{t}^{\mathrm{DC} }=A_{ t}^{\mathrm{IFL} }+A_{ t}^{\mathrm{FL} },\forall t \in T \label{dc3}
\end{align}
where $P^{\text{idle}}$ and $P^{\text{peak}}$ denote the power consumption of a server when it is not processing tasks and when it is actively processing tasks, respectively. $W_t^{\text{DC}}$ denotes the total arriving workloads in the data center, $L_{j,t}^{\text{FL}}$ denotes the arriving class $c$ workloads at time $t$, and $L_{t}^{\text{IFL}}$ represents the arriving inflexible workloads at time $t$. $A_t^{\text{IFL}}$ and $A_t^{\text{FL}}$ are the number of active servers for inflexible workloads and flexible workloads, respectively. $L^{\text{rate}}$ is the service rate of a server. Eq. (\ref{dc2}) calculates the total workload, including both flexible and inflexible workloads.  Eq. (\ref{dc3}) indicates the number of active servers in the data center.

\subsection{Data Centers Workloads Scheduling Model}
\textbf{Flexible workloads.}
Flexible workloads, such as big data analysis, image processing, and AI training, can be deferred within a reasonable time. Following the model in \cite{hall2024carbon}, it is assumed that each class of workloads $c \in \mathcal{C}$ is associated with a temporal flexibility parameter $h_c \in \mathbb{Z}_{\geq 0}$, which represents the maximum allowable delay for workloads of that class. Then the flexible workloads scheduling model is as follows:
\begin{align}
& \sum\nolimits_{\tau=1}^t L_{C, \tau}^{\mathrm{FL}} \leq \sum\nolimits_{\tau=1}^t \bar{L}_{C, \tau}^{\mathrm{FL}},  \forall t \in \mathcal{T}, \forall c \in \mathcal{C}, \label{fl1}\\
& \sum\nolimits_{\tau=1}^{t+h_c} L_{c, \tau}^{\mathrm{FL} } \geq \sum\nolimits_{\tau=1}^t \bar{L}_{c, \tau}^{\mathrm{FL}},  \forall t \in\left[1,|T|-h_c\right],\forall c \in \mathcal{C},\label{fl2}\\
& \sum\nolimits_{\tau=1}^T L_{c, \tau}^{\mathrm{FL} } \geq \sum\nolimits_{\tau=1}^t \bar{L}_{ c,\tau}^{\mathrm{FL}},  \forall t \in\left[|T|-h_c+1,|T|\right],\forall c \in \mathcal{C},\label{fl3}\\
& L^{\mathrm{rate}} A_{ t}^{\mathrm{FL} } \geq \sum\nolimits_{c \in \mathcal{C}} L_{c, t}^{\mathrm{FL} }, \forall t \in \mathcal{T} \label{fl4}\\
& L_{t}^{\mathrm{FL}}, A_{t}^{\mathrm{FL} } \geq 0, \forall t \in \mathcal{T} . \label{fl5}
\end{align}
Eq. (\ref{fl1}) indicates that the processed flexible workload does not exceed the accumulated flexible workload up to time $t$. Eq. (\ref{fl2}) specifies that the accumulated flexible workload must be processed within the maximum allowable delay time $h_c$.
Eq. (\ref{fl3}) suggests that all flexible workloads should be processed within a single day. It is assumed that all the flexible workloads processed in the homogeneous active servers must satisfy the total number of inflexible workloads as specified in Eq. (\ref{fl4}). Eq. (\ref{fl5}) ensures that the number of inflexible workloads and the corresponding active servers are positive.

\textbf{Inflexible workloads.} Within data centers, there exist user-facing services, such as inference, which mandate real-time responses and require immediate processing. These are categorized as inflexible workloads, and their operation model is expressed as follows:
\begin{align}
 &L_{t}^{\mathrm{IFL}} \geq \bar{L}_t^{\mathrm{IFL}}, \forall t \in \mathcal{T},\label{in1}\\
 & \frac{1}{L_j^{\mathrm{rate}}-L_{j, t}^{\mathrm{IFL}} / A_{j, t}^{\mathrm{IFL} }} \leq \mathrm{C}^{\mathrm{DT}}, \forall j \in D, \forall t \in \mathcal{T},\label{in2}\\
& L_{j, t}^{\mathrm{IFL}}, A_{j, t}^{\mathrm{IFL}} \geq 0, \forall j \in D, \forall t \in \mathcal{T}, \label{in3}\\
& A_{j, t}^{\mathrm{DC}} \leq A_{j, \max }^{\mathrm{DC}}, \forall j \in D, \forall t \in \mathcal{T},  \label{in4}
\end{align}
Eq. (\ref{in1}) indicates that the inflexible workloads dispatched in the data center must meet or exceed the inflexible workloads arriving in the front-end server.  Eq. (\ref{in2}) describes that active servers responsible for processing inflexible workloads must comply with user time-delay requirements, as defined by the quality of service (QoS).  Eq. (\ref{in3}) ensures that both the number of inflexible workloads and the corresponding active servers are positive. Eq. (\ref{in4}) restricts that the number of active servers in the first stage should not exceed the total number of servers.

2) \textit{Grid Carbon-based Energy Accounting}

This study assumes that the data center is connected to the main power grid. The power supply of the grid is derived from a mix of energy sources, including carbon-free energy such as wind, solar, and nuclear energy, as well as carbon-based energy such as oil, natural gas, and coal-fired power. To allow for a more accurate assessment, a carbon-based energy accounting model is employed.  This model quantifies the proportion of carbon-based energy per unit of electricity generated, as defined by the following equation:
\begin{align}
\sum\nolimits_{t=1}^{T} CB_t \cdot P_{buy,t}, \label{eq:cb}
\end{align}
where $P_{buy,t}$ is the power purchased by the data center at time $t$, and $CB_t$ denotes the proportion of carbon-based energy in the grid at time $t$. It is further expressed as:
\begin{align}
CB_t = \frac{\sum_{k\in{\mathcal{E}_{CB}}} M^k_t}{\sum_{k \in \mathcal{E}} M^k_t},
\label{eq:ci}
\end{align}
where $\mathcal{E}$ represents the set of all energy sources (e.g., wind, solar, coal), $ \mathcal{E}_{CB}$ denotes the subset of carbon-based energy sources (e.g., coal, oil, natural gas), and $M^k_t$ is the electricity output from energy source $k$ at time $t$. 

$CB_t$ reflects the average proportion of carbon-based energy in the grid at a given time and varies hourly depending on the real-time contribution of each energy source. It provides a more accurate measurement of carbon emissions associated with electricity consumption. And real-time data on the composition of energy sources is usually provided by Independent System Operators (ISOs) \cite{CISO}. To achieve 24/7 CFE, data centers are required to acquire Time-Based Energy Attribute Certificates (T-EACs). The total T-EACs purchased by the data center must correspond to the carbon-based energy consumed by the data center, as defined in Eq. (\ref{eq:cb}).

Traditional Energy Attribute Certificates (EACs) match renewable generation with consumption annually, allowing claims based on total certificates purchased, irrespective of generation timing. In contrast, time-based EACs enforce stricter hourly or real-time matching, ensuring that renewable energy is claimed only when generated concurrently with consumption.
This granular T-EAC approach improves the accuracy of renewable usage claims and incentivizes investments in energy storage and grid management to better synchronize supply and demand, supporting 24/7 carbon-free energy systems. 



\subsection{ESS Model}
As energy storage technology has advanced, energy storage systems (ESS) have become a widely adopted and cost-effective solution for storing renewable energy \cite{acun2023carbon}. Data centers usually incorporate ESS to enhance system resilience and mitigate power peaks. Moreover, ESS can support 24/7 CFE operations by storing surplus power when renewable energy generation exceeds demand and discharging stored energy during periods of insufficient renewable supply.
The ESS is modeled as:
\begin{align}
& Q_{ess,t} = Q_{ess,t-1} + P_{ess,t}, \label{bies:1}  \\
& Q_{ess,min }\leq Q_{ess,t} \leq Q_{ess,max}, \label{bies:3}\\
& P_{ess,min }\leq P_{ess,t} \leq P_{ess,max}, \label{bies:2}
\end{align}
where $P_{ess,min }$ and $P_{ess,max}$ are the minimum and maximum charge and discharge power of the ESS. $P_{ess,t}$ is the charging and discharging power of ESS at time $t$ (positive when charging and negative when discharging). $Q_{ess,t}$ denotes the level of electricity storage of the ESS
at time $t$.

\subsection{Chance-Constraints Problem Formulation}

For the data center, the power supply must be greater than the power demand. Considering the uncertainty in renewable energy generation, the power balance for electricity is expressed as chance constraints, formulated as follows:
\begin{align}
&\operatorname{Pr}\left(P_{t}^{\text{renew}}+P_{t}^{\text{e,grid}} \geq  Q_{t}^{DC} +P_t^{ESS} \right) \geq 1 -\epsilon, \forall t \in \mathcal{T},  \label{cc:1}
\end{align}
where $P_{t}^{\text{e,grid}}$ is the purchased electricity at time $t$; $P_{t}^{\text{renew}}$ is the renewable energy generation at time $t$. $\epsilon$ is the violation rate. 

Therefore, the overall 24/7 CEF data centers operation scheduling problem is formulated as follows:
\begin{align}
\rm{(CCP)} \min &\sum\nolimits_{t=1}^T \lambda_t^{e} P_{t}^{\text{grid}}+\lambda^{c}CE,\\
&s.t. (\ref{dc1}) - (\ref{cc:1}).
\end{align}
where $\lambda_t^{\text{e}}$ and $\lambda^{\text{c}}$ are the electricity price and T-EAC price, respectively.  The objective of the data center microgrid is to minimize the total operational cost, which comprises the costs of energy procurement and the costs associated with purchasing T-EACs.

\section{Proposed Framework of Conformal Prediction-based Chance Constrained Programming}
\label{robust_optimization}

In this section, we begin by revisiting the formulation of CCP and present the fundamental procedural framework of our algorithm. Subsequently, we present the methodologies that have been developed in our research.

\subsection{Contextual Chance-Constraints Programming}

The common form of a traditional CCP typically takes the following structure:
\begin{align}
\text{(T-CCP) min} \ f(x); \ \text{s.t.} \ P(g(x ; \xi) \in \mathcal{A}) \geq 1-\epsilon,
\label{m1}
\end{align}
where $x$ is the decision variable, $\xi$ is a random variable that can be observed via a finite amount of data, $\mathcal{A}$ is the feasible region, and $\epsilon$ is the tolerance level.

In the traditional CCP, the random variable 
$\xi$  is estimated using a finite set of historical data. However, this estimation process is independent of any additional features or information, which limits its modeling capability when dealing with complex distributions in practical scenarios. With the increasing availability of data, there has been a growing interest in formulating optimization under uncertainty as contextual optimization problems. These problems aim to leverage rich feature observations to improve decision-making \cite{ban2019big}. In this context, we employ a contextual CCP (C-CCP) approach \cite{rahimian2020contextual}:
\begin{align}
\text{(C-CCP) min} \ f(x); \ \text{s.t.} \ P_{\xi \mid \psi}(g(x ; \xi) \in \mathcal{A}) \geq 1-\epsilon.
\label{model:c-ccp}
\end{align}

The probability is considered with respect to the conditional distribution ${\xi \mid \psi}$. The "contextual" decision maker has access to a vector of covariates $z$,  which is assumed to be correlated with $\xi$. Consequently, this decision maker seeks to identify an optimal policy, defined as a functional  $\boldsymbol{x}: \mathbb{R}^m \rightarrow \mathcal{X}$ that gets an action in $\mathcal{X}$ based on the observed realization of 
$\psi$  \cite{chenreddy2022data}.


\subsection{The RO Reformulation of the C-CCP}
\label{Reformulation}

The original chance constraints are intractable due to their non-differentiability and non-convexity, necessitating the use of an approximation. To address this problem, the model described in (\ref{m1}) is transformed into a robust optimization (RO) problem. This transformation also enables the direct utilization of data \cite{bertsimas2018data}. Robust optimization addresses uncertainty by representing it through a deterministic set, known as an uncertainty set. The RO formulation ensures that the safety condition is satisfied for any value of 
$\xi$  within the uncertainty set, thereby simplifying the problem and enhancing its tractability. The RO formulation of (\ref{m1}) is as follows:
\begin{align}
\text {(RO) min } f(x) \text { s.t. } g(x ; \xi) \in \mathcal{A}, \forall \xi \in \mathcal{U}, \label{model:ro1}
\end{align}
 where  $\mathcal{U} \in \Omega $ is an uncertainty set. The theorem for the reformulation is presented as follows:
\begin{theorem}
   Any feasible solution to problem RO (\ref{model:ro1}) is also feasible to problem T-CCP (\ref{m1}) if the following condition holds for $\mathcal{U}$:
    \begin{align}
    \mathbb{P}_{\xi}(\boldsymbol{\xi} \in \mathcal{U}) \geq \rho
    \end{align}
\end{theorem}
\begin{proof}
 Obviously, for any $x$ feasible for (\ref{model:ro}), $\xi \in \mathcal{U}$ implies $g(x ; \xi) \in \mathcal{A}$. Therefore, by choosing $\mathcal{U}$ that covers a $1-\epsilon$ content of $\xi$ (i.e., $\mathcal{U}$ satisfies $P(\xi \in \mathcal{U}) \geq 1-\epsilon$ ), any $x$ feasible for (\ref{model:ro}) must satisfy $P(g(x ; \xi) \in \mathcal{A}) \geq P(\xi \in \mathcal{U}) \geq 1-\epsilon$, implying that $x$ is also feasible for (\ref{m1}). 
\end{proof}
Therefore, we can address the contextual chance-constrained problem from the perspective of robust optimization. Specifically, we consider a contextual decision maker who aims to utilize side information in the design and solution of a robust optimization problem. This approach naturally leads to the following equivalent formulation of C-CCP:
\begin{align}
\text { min } f(x) \text { s.t. } g(x ; \xi) \in \mathcal{A}, \forall \xi \in \mathcal{U}(\psi) \label{model:ro}
\end{align}
where $\mathcal{U}(\psi)$ represents an uncertainty set that is constructed to encompass, with high probability, the realization of $\xi$ given the observation of $\psi$. The proposed approach is data-driven, as it utilizes historical observations of joint realizations of $\psi$ and $\xi$ in the design of the contextual robust optimization problem.

Following the schedule outlined in Theorem 1, we can conclude the following:
\begin{theorem}
    Any feasible solution to the robust optimization problem RO (\ref{model:ro}) is also feasible for the contextual chance-constrained problem C-CCP (\ref{model:c-ccp}), if the following condition holds for $\mathcal{U}$:
    \begin{align}
    \mathbb{P}_{\xi \mid \psi}(\boldsymbol{\xi} \in \mathcal{U}(\psi)) \geq \rho.
    \end{align}
\end{theorem}
This leads to the challenge of determining the uncertainty set when side information 
$\psi$ is available. To address this issue, we introduce the idea of uncertainty quantification and conformal prediction.
\subsection{Uncertainty Quantification and Conformal Prediction}
Conformal prediction is a statistical and distribution-free technique that quantifies uncertainty for any black-box predictive model. It was first introduced by Vork~\cite{vovk2005algorithmic} as a postprocessing method and can be applicable for both classification and regression tasks. Consider a dataset $\mathcal{D}=\{(X_i, Y_i)\}_{i=1}^{n}$, where each data point $(X_i, Y_i)$ comprises a covariate vector $X_i$ and a corresponding response $Y_i$. Conformal prediction aims to find a prediction band $C(X_{n+1})$ for a new observation $X_{n+1}$. This prediction band is designed to contain the true, but unknown response $Y_{n+1}$ with a user-specified probability under the assumption that $\mathcal{D}$ and $(X_{n+1}, Y_{n+1})$ are exchangeable \cite{lei2018distribution}. Specifically, for a significance level $\alpha$, we aim to construct $C(X_{n+1})$ such that:
\begin{equation}\label{eq:cp-lb1}
\mathbb{P}(Y_{n+1} \in C(X_{n+1})) \geq 1 - \alpha,
\end{equation}
Building on the foundation of conformal prediction, we now focus on the practical applications in real-world scenarios.  The full conformal prediction is computationally expensive as the construction of the prediction set must invert the score function to identify possible values $y \in \mathcal{Y}$ for the response $Y_{n+1}$ that agree (or conform) with the trends observed in the available data~\cite{angelopoulos2024theoretical}. 

To address this, split conformal prediction was developed to offer a computationally efficient alternative by dividing the data to avoid the need to retrain the conformal score function. Although our proposed approaches can apply to both versions of conformal prediction,  we restrict our attention to the split conformal prediction and refer the reader to~\cite{angelopoulos2024theoretical} for a more detailed comparison between these two methods.

Split conformal prediction begins by partitioning the training data $\mathcal{D}$ into two disjoint sets: a proper training set $\mathcal{I}_1$ and a calibration set $\mathcal{I}_2$.  First, we train a regression model $\hat{\mu}(x)$ on the proper training set using any regression algorithm $\mathcal{A}$:
\begin{align}
\hat{\mu}(x) \leftarrow \mathcal{A}\left(\left\{\left(X_i, Y_i\right): i \in \mathcal{I}_1\right\}\right) .\label{eq:cp1}
\end{align}

Next, we calculate the calibration residuals $R_i$ on the calibration set, which measure the prediction error:
\begin{align}
R_i=\left|Y_i-\hat{\mu}\left(X_i\right)\right|, \quad i \in \mathcal{I}_2 .\label{eq:cp-residual}
\end{align}

To determine the width of our prediction band, we compute the $(1-\alpha)$-quantile of these residuals. Specifically, we use a slightly conservative quantile $Q_{1-\alpha}\left(R, \mathcal{I}_2\right)$ to ensure coverage:
\begin{align}
Q_{1-\alpha}:=quantile(\{R_i: i \in \mathcal{I}_2\}, (1-\alpha)\left(1+1 /\left|\mathcal{I}_2\right|\right)). 
\end{align}

Finally, for a new data point $X_{n+1}$, the split conformal prediction interval is constructed as:
\begin{align}
C\left(X_{n+1}\right)=[\hat{\mu}\left(X_{n+1}\right)-Q_{1-\alpha}, \ \hat{\mu}\left(X_{n+1}\right)+Q_{1-\alpha}] .
\end{align}
This interval is guaranteed to satisfy the coverage property in equation~\eqref{eq:cp-lb1}. In summary, we have the following Theorem:
\begin{theorem}\label{theorem:theorem3}
If dataset ${\mathcal{D}=(X_i, Y_i)}$ for any $i \in \{1, …, n\}$ are exchangeable. For a new exchangeable pair $(X_{n+1}, Y_{n+1})$, conformal prediction construct a prediction set $C(X_{n+1})$ with the following marginal coverage guarantee:
\begin{align} 
    \label{eq: cp-lower-bound}
    \mathbb{P}(Y_{n+1} \in C(X_{n+1})) \geq 1-\alpha,
\end{align}
where $\alpha$ is a predefined confidence level. Note that if $R_i$ has a continuous joint distribution, then equation \eqref{eq: cp-lower-bound} has a tigher bound:
\begin{align} 
    \label{eq: cp-upper-bound}
    \mathbb{P}(Y_{n+1} \in C(X_{n+1})) \leq 1-\alpha + \frac{1}{n+1},
\end{align}

\begin{proof}[Proof for lower bound]
Obvious, the lower bound $\mathbb{P}(Y_{n+1} \in C(X_{n+1})) = P(R_{n+1} \leq Q_{1 - \alpha}) \geq 1 - \alpha$ holds because $Q_{1-\alpha}$ is defined as the $(1-\alpha)$ quantile of calibration residuals.  By definition, this quantile ensures that at least a $(1-\alpha)$ proportion of calibration residuals (and thus, in expectation, future residuals) are smaller, guaranteeing the desired coverage.
\end{proof}
\begin{proof}[Proof for upper bound]
If $R_i$ have a continuous joint distribution, then with probability 1, all scores are distinct.  In this case, assume $k = \lceil (1-\alpha)(n+1) \rceil$, then $\mathbb{P}(Y_{n+1} \in C(X_{n+1})) = P(R_{n+1} \leq Q_{1 - \alpha}) = P(R_{n+1} \leq S_k) = \frac{k}{n+1} = \frac{\lceil (1-\alpha)(n+1) \rceil}{n+1} \leq 1-\alpha + \frac{1}{n+1}$.
\end{proof}
\end{theorem}
Therefore, for a new data sample $X_{n+1}$, the interval $Y_{n+1}$ can be interpreted as the uncertainty set $\mathcal{U}(X_{n+1})$ described in Section \ref{Reformulation}. Consequently, the problem can be addressed using conventional robust optimization techniques \cite{ben2009robust}. Note that in this research, we consider the box-uncertainty set. It can also be extended to the elliptical-uncertainty set, we refer \cite{xu2024conformal} for more details.  A key limitation of this basic split conformal interval is that it applies a uniform interval \( 2Q_{1-\alpha} \) to all the samples without accounting for the varying difficulty of each sample. This motivates the need for more adaptive approaches, such as variable-width conformal prediction methods.

\section{The Proposed Approach}
\label{conformal_prediction}
\subsection{Conformal Quantile Regression}

Unlike the uniform width interval of standard split conformal prediction, conformal quantile regression (CQR) leverages separately trained lower and upper quantile predictions to adapt the interval width based on the input \(X_{n+1}\).  This approach aims to provide more informative and efficient prediction intervals while maintaining the same statistical coverage property. Fig. \ref{fig:cp} illustrate the process of getting \(X_{n+1}\).

Suppose $\mu(X; \hat{\theta}_{l})$ and $\mu(X; \hat{\theta}_{h})$ be the lower and upper quantile regression models trained on the proper training set $\mathcal{I}_1$. Compared with Eq.~\eqref{eq:cp-residual}, the CQR-based method builds the calibration residuals $R_i^l$ and $R_i^h$ by lower bound and upper bound separately. Let’s define the nonconformity scores on the calibration set $\mathcal{I}_2$ as:
\begin{align}
\label{eq: cqr-l1}
R_i^{l} = \mu(X_i;\hat{\theta}_{l}) - Y_i, \quad i \in \mathcal{I}_2, \\
R_i^{h}  = Y_i - \mu(X_i;\hat{\theta}_{h}), \quad i \in \mathcal{I}_2.
\end{align}
Let $Q_{1-\alpha_{l}}$ be the $(1-\alpha_{l})$-th quantile of $\{R_i^{l}: i \in \mathcal{I}_2\}$ adjusted with the factor $(1+1/|\mathcal{I}_2|)$, and $Q_{1-\alpha_{h}}$ be the $(1-\alpha_{h})$-th quantile of $\{R_i^{h}: i \in \mathcal{I}_2\}$ adjusted with the factor $(1+1/|\mathcal{I}_2|)$. Construct the prediction interval for a new data point $X_{n+1}$ as:
\begin{equation*}
C(X_{n+1}) = \left[ \mu(X_{n+1}; \hat{\theta}_{l}) - Q_{1-\alpha_{l}},  \mu(X_{n+1}; \hat{\theta}_{h}) + Q_{1-\alpha_{h}}   \right].
\end{equation*}
Then the following theorem can be derived:

\begin{figure}[t]
\centering
\includegraphics[width=0.7\columnwidth]{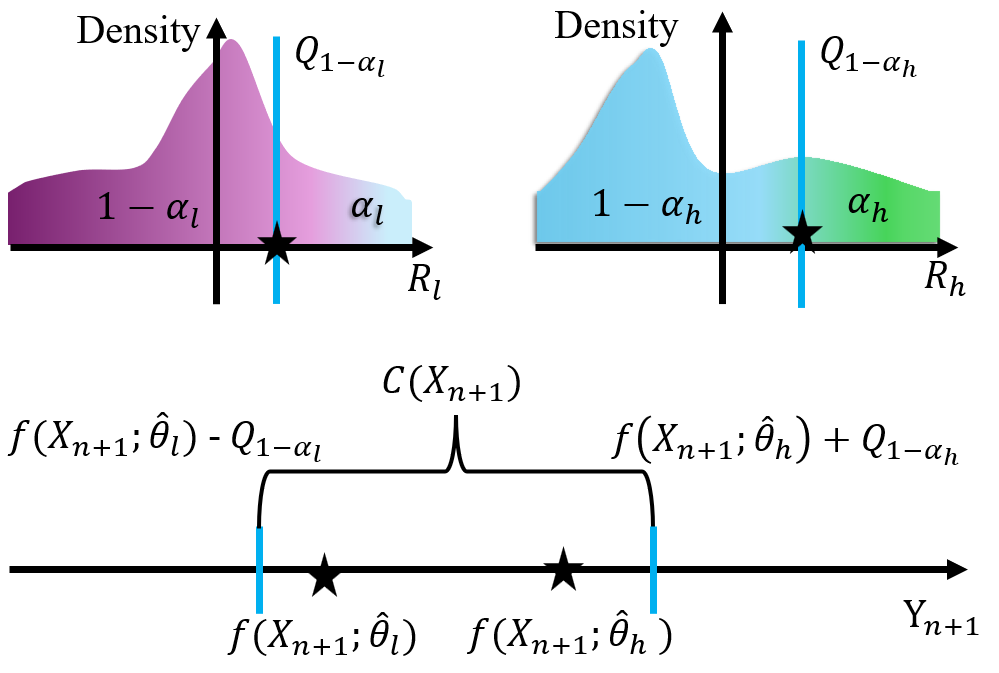}
\caption{The process getting  prediction interval $C(X_{n+1})$.}
\label{fig:cp}
\end{figure}

\begin{theorem}
For a given significance level $\alpha = \alpha_{l} + \alpha_{h}$, if dataset ${\mathcal{D}=(X_i, Y_i)}$ for any $i \in \{1, …, n\}$ are exchangeable, the CQR prediction interval for the new data point pair $X_{n+1}$ satisfies the following marginal coverage guarantee:
\begin{align} 
    \label{eq: cqr-lower-bound}
    \mathbb{P}(Y_{n+1} \in C(X_{n+1})) \geq 1- \alpha.
\end{align}
\vspace{-10mm}
\begin{proof}
According to the construction process of the prediction intervals, the lower quantile residual \( Q_{1-\alpha_l} \) and the upper quantile residual \( Q_{1-\alpha_h} \) satisfy the following properties:
\begin{align}
\mathbb{P}\big(Y_{n+1} \geq \mu(X_{n+1}; \hat{\theta}_l) - Q_{1-\alpha_l}\big) \geq 1 - \alpha_l, \\
\mathbb{P}\big(Y_{n+1} \leq \mu(X_{n+1}; \hat{\theta}_h) + Q_{1-\alpha_h}\big) \geq 1 - \alpha_h.
\end{align}
Then prediction interval \( C(X_{n+1}) \) covers \( Y_{n+1} \) if both inequalities hold. Let \( A = \{Y_{n+1} \geq \mu_l - Q_{1-\alpha_l}\} \) and \( B = \{Y_{n+1} \leq \mu_h + Q_{1-\alpha_h}\} \). Using the union bound on their complements, we can get:
\begin{align}
\mathbb{P}(A \cap B) &= 1 - \mathbb{P}(A^c \cup B^c) \\  &\geq 1 - (\mathbb{P}(A^c) + \mathbb{P}(B^c)) \\
&\geq 1 - (\alpha_l + \alpha_h).
\end{align}
\end{proof}
\end{theorem}
\vspace{-6mm}
The CQR method achieves valid marginal coverage guarantees while adapting interval widths to input-dependent uncertainty through quantile regression. Leveraging data-dependent adjustments, it produces narrower intervals in regions of low uncertainty and wider intervals where variability is higher. This adaptability often leads to tighter, more informative prediction intervals without sacrificing coverage.
\subsection{Multi-Variable Conformal Quantile Prediction}

In many real-world applications, the target response variable $Y$ is not a scalar but a vector, i.e., $Y \in \mathbb{R}^d$. In this section, we extend conformal prediction to multi-variable outputs and demonstrate two distinct approaches to multi-variable conformal quantile prediction: one focusing on marginal coverage across all output variables and another aiming for individual coverage for each output variable.

\subsubsection{Average Marginal Multi-Variable CQR}

In this section, we consider a weaker notion of marginal coverage for multi-variable output, which we term Average Marginal Multi-Variable CQR (AMV-CQR). Here, we aim to ensure that the average coverage across all output dimensions is at least $1-\alpha$. Formally, we want to satisfy:
\begin{equation}\label{eq:mv-average-marginal-coverage}
\frac{1}{d} \sum_{j=1}^{d} \mathbb{P}(Y_{n+1}^j \in C^j(X_{n+1})) \geq 1 - \alpha.
\end{equation}
where $Y_{n+1} = [Y_{n+1}^1, \ldots, Y_{n+1}^d]^\top$ and $Y_{n+1}^j$ is the j-th output variable. $C(X_{n+1}) = C^1(X_{n+1}) \times \ldots \times C^d(X_{n+1})$ is a hyper-rectangular prediction region, with $C^j(X_{n+1})$ being the prediction interval for the $j$-th output variable.

Note that we need a multivariate output model for AMV-CQR, which can generate lower and upper bounds for all response variables. For brevity, we share the lower and upper bound models for multivariate prediction and only change the heads of different outputs. The details of the process are described in Algorithm \ref{alg:ammv-cqr}. The following theorem establishes that the intervals obtained through Algorithm~\ref{alg:ammv-cqr} satisfy the marginal coverage guarantee:

\begin{algorithm}[t]
\caption{Average Marginal Multi-Variable CQR (AMV-CQR)}
\label{alg:ammv-cqr}
\small
\begin{algorithmic}[1]
    \Require $\mathcal{D}=\{(X_i, Y_i)\}_{i=1}^{n}$, $Y_i \in \mathbb{R}^d$, significance $\alpha$, quantiles $\alpha_{l}$, $\alpha_{h}$, where $\alpha_{l} + \alpha_{h} = \alpha$, the index set $\mathcal{J}=\{1, 2, \ldots, d\}$.
    \Ensure Prediction intervals $C^1(X_{n+1}), \ldots, C^d(X_{n+1})$.
    \State Split $\mathcal{D}$ into proper training set $\mathcal{I}_1$ and calibration set $\mathcal{I}_2$.
    \State Train lower and upper quantile models $\mu(\cdot;\theta_l), \mu(\cdot;\theta_h)$ on proper training set $\mathcal{I}_1$.
    \For{ $j \in \mathcal{J}, i \in \mathcal{I}_2$} \State Calculate residuals: $$R_i^{l, j} = \mu^j(X_i;\theta_l) - Y_i^j, R_i^{h, j} = Y_i^j - \mu^j(X_i;\theta_l).$$
    \State Compute quantiles:
    $$ \beta_l=(1-\alpha_l)(1+1/{(|\mathcal{I}_2|d)}), 
        \beta_h=(1-\alpha_h)(1+1/{(|\mathcal{I}_2|d)}),$$
    \begin{align*}
        Q_{1-\alpha_l} &= quantile\left(\{R_i^{l,j} : i \in \mathcal{I}_2,\; j \in \mathcal{J}\},\; \beta_l \right), \\
        Q_{1-\alpha_h} &= quantile\left(\{R_i^{h,j} : i \in \mathcal{I}_2,\; j \in \mathcal{J}\},\; \beta_h \right).
    \end{align*}
    \EndFor
    \For{ $j \in \mathcal{J}$} 
    \State Construct prediction interval $X_{n+1}^j$:
    \begin{align*}
        C^j(X_{n+1}) = \left[ \mu^j(X_{n+1};\theta_l) - Q_{1-\alpha_l},  \mu^j(X_{n+1};\theta_h) + Q_{1-\alpha_h}   \right]
    \end{align*}
    \EndFor
    \State Return the prediction intervals $C^1(X_{n+1}), \ldots, C^d(X_{n+1})$.
\end{algorithmic}
\end{algorithm}

\begin{theorem}
\label{thm:ammv-cqr-coverage}
If dataset ${\mathcal{D}=(X_i, Y_i)}$ for any $i \in \{1, \ldots, n\}$ are exchangeable, and $Y_i \in \mathbb{R}^d$. For a new exchangeable pair $(X_{n+1}, Y_{n+1})$, the AMV-CQR algorithm constructs prediction intervals $C^1(X_{n+1}), \ldots, C^d(X_{n+1})$ such that the average marginal coverage guarantee is satisfied:
\begin{align*}
    \frac{1}{d} \sum_{j=1}^{d} \mathbb{P}(Y_{n+1}^j \in C^j(X_{n+1})) \geq 1 - \alpha.
\end{align*}
\end{theorem}

\begin{proof}
The result follows from a simple extension of the univariate case. Since the calibration residuals across both indices $j \in \mathcal{J}$ are exchangeable, we can flatten them and apply the standard conformal procedure. Hence, as in Theorem~\ref{theorem:theorem3}, the lower bound and upper bound of the prediction set hold. 
\end{proof}

\subsection{Individual Multi-Variable CQR}

In contrast to average marginal coverage, individual coverage focuses on ensuring coverage for each output variable separately. For a given significance level $\alpha$, we aim to construct prediction intervals $C^j(X_{n+1})$ for each output dimension $j \in \{1, \ldots, d\}$ such that for each $j$:
\begin{equation}\label{eq:mv-individual-coverage}
\mathbb{P}(Y_{n+1}^j \in C^j(X_{n+1})) \geq 1 - \alpha.
\end{equation}
Note that individual coverage is a stronger guarantee than average marginal coverage. If individual coverage holds, average marginal coverage automatically holds for the same $\alpha$. In this section, we propose the Individual Multi-Variable CQR (IMV-CQR) algorithm to achieve individual coverage. The specifics of the process are outlined in Algorithm 2.  The following theorem establishes that the intervals obtained through Algorithm~2 satisfy the individual coverage guarantee:

\begin{algorithm}[t]
\begin{small}\caption{Individual Multi-Variable Conformal Quantile Regression (IMV-CQR)}
\end{small}
\label{alg:imvcqr}
\begin{small}
\begin{algorithmic}[1]
    \Require Training data $\mathcal{D}=\{(X_i, Y_i)\}_{i=1}^{n}$ where $Y_i \in \mathbb{R}^d$, significance $\alpha$, quantiles $\alpha_{l}$, $\alpha_{h}$, where $\alpha_{l} + \alpha_{h} = \alpha$, the index set $\mathcal{J}=\{1, 2, \ldots, d\}$. New data point $X_{n+1}$.
    \Ensure Prediction intervals $C^1(X_{n+1}), \ldots, C^d(X_{n+1})$.
    \State Split $\mathcal{D}$ into proper training set $\mathcal{I}_1$ and calibration set $\mathcal{I}_2$.
    \State Train lower and upper quantile models $\mu(\cdot;\theta_l), \mu(\cdot;\theta_h)$ on proper training set $\mathcal{I}_1$.
    \For{ $j \in \mathcal{J}, i \in \mathcal{I}_2$}
    \State Calculate residuals:
    $$R_i^{l, j} = \mu^j(X_i;\theta_l) - Y_i^j ,
        R_i^{h, j} = Y_i^j - \mu^j(X_i;\theta_l).$$
    \EndFor
    \For{$j \in \mathcal{J}$}
    \State Compute quantiles:
    $$    \beta_l=(1-\alpha_l)(1+1/{|\mathcal{I}_2|}) ,
        \beta_h=(1-\alpha_h)(1+1/{|\mathcal{I}_2|}) $$
    \begin{align*}
        Q_{1-\alpha_l}^{j} &= quantile\left(\{R_i^{l,j} : i \in \mathcal{I}_2\},\; \beta_l\right), \\
        Q_{1-\alpha_h}^{j} &= quantile\left(\{R_i^{h,j} : i \in \mathcal{I}_2\;\},\; \beta_h\right).
    \end{align*}
    \EndFor
    \For{$j \in \mathcal{J}$}  \State Construct prediction interval for $X_{n+1}^j$:
    \begin{align*}
        C^j(X_{n+1}) = \left[ \mu^j(X_{n+1}; \hat{\theta}_{l}) - Q_{1-\alpha_l}^{j},  \mu^j(X_{n+1}; \hat{\theta}_{h}) + Q_{1-\alpha_h}^{j}   \right]
    \end{align*}
    \EndFor
    \State Return the  prediction intervals $C^1(X_{n+1}), \ldots, C^d(X_{n+1})$.
\end{algorithmic}
\end{small}
\end{algorithm}

\begin{theorem}
\label{thm:imv-cqr-coverage}
If dataset ${\mathcal{D}=(X_i, Y_i)}$ for any $i \in \{1, \ldots, n\}$ are exchangeable, and $Y_i \in \mathbb{R}^d$. For a new exchangeable pair $(X_{n+1}, Y_{n+1})$, the IMV-CQR algorithm constructs prediction intervals $C^1(X_{n+1}), \ldots, C^d(X_{n+1})$ such that for each dimension $j \in \{1, \ldots, d\}$, the individual coverage guarantee is satisfied:
\begin{align}
    \mathbb{P}(Y_{n+1}^j \in C^j(X_{n+1})) \geq 1 - \alpha.
\end{align}

\begin{proof}
The result also follows from a simple extension of the univariate case. Since the calibration residuals across both indices $j \in \mathcal{J}$ are exchangeable, we apply the standard conformal procedure to each output separately. Hence, as in Theorem~\ref{theorem:theorem3}, the lower bound and upper bound of the prediction set hold. 
\end{proof}
\end{theorem}
\begin{figure}[t]
\centering
\includegraphics[width=1\columnwidth]{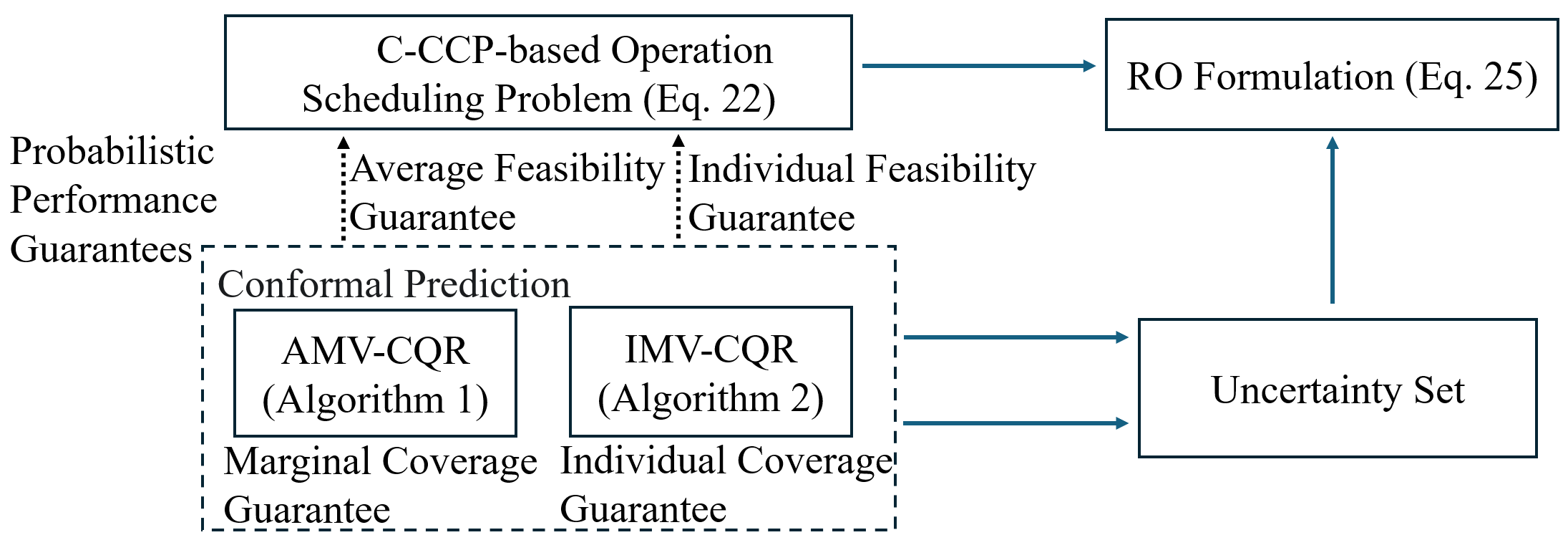}
\caption{The overall process}
\label{fig:process}

\end{figure}
\subsection{Comparing AMV-CQR versus IMV-CQR}

AMV-CQR targets a collective or average marginal coverage guarantee, which is formalized as follows:
\begin{equation}\label{eq:mv-average-marginal-coverage}
\frac{1}{d} \sum_{j=1}^{d} \mathbb{P}\Big(Y_{n+1}^j \in C^j(X_{n+1})\Big) \geq 1 - \alpha.
\end{equation}
In practice, this means that across all $d$ output variables, the mean coverage meets the desired level $1-\alpha$, even if some variables may be slightly undercovered, provided that others are overcovered. This relaxed requirement may yield tighter overall prediction sets and is well-suited in applications where only the aggregate reliability is of interest or when the outputs exhibit strong correlations.

In contrast, IMV-CQR enforces an individual coverage guarantee for each component $j \in \{1,\ldots,d\}$, as described by
\begin{equation}\label{eq:mv-individual-coverage}
\mathbb{P}\Big(Y_{n+1}^j \in C^j(X_{n+1})\Big) \geq 1 - \alpha.
\end{equation}
It is inherently stronger and more desirable when each individual prediction is required to meet a strict reliability standard. However, the additional calibration burden makes IMV-CQR more conservative in practice. Because calibration is performed for each output channel separately, the method may yield wider intervals, especially when the sample size is limited. 

Building upon Theorem 2 and leveraging Eqs. (\ref{eq:mv-average-marginal-coverage}) and (\ref{eq:mv-individual-coverage}), we establish that the uncertainty sets generated by our conformal prediction-based methods directly translate to statistical feasibility guarantees for the original C-CCP problem when used within the reformulated problem (\ref{model:ro}). This implies that using uncertainty sets from AMV-CQR yields an average statistical feasibility guarantee for C-CCP, while uncertainty sets derived from IMV-CQR ensure an individual statistical feasibility guarantee.  Fig. \ref{fig:process} provides a summary of this overall methodology and the relationships between different components.

\section{Numerical studies}

\label{Numerical_studies}
\subsection{Data Description and Experimental Settings}

In this section, comprehensive numerical studies are conducted to validate the performance of the proposed conformal prediction-based chance-constraints algorithms and to analyze the sensitivity of key parameters. The parameters required for the components in the datacenter are listed in Table 1, which are set based on the research papers by \cite{yang2023two} and \cite{zhou2022bi}. All simulations were implemented using Python on a desktop computer equipped with an Intel® i9 2.4GHz CPU and 32GB RAM. The optimization problem is solved using Gurobi 11.0.3.

\begin{figure}[t]
\centering
\includegraphics[width=0.7\columnwidth]{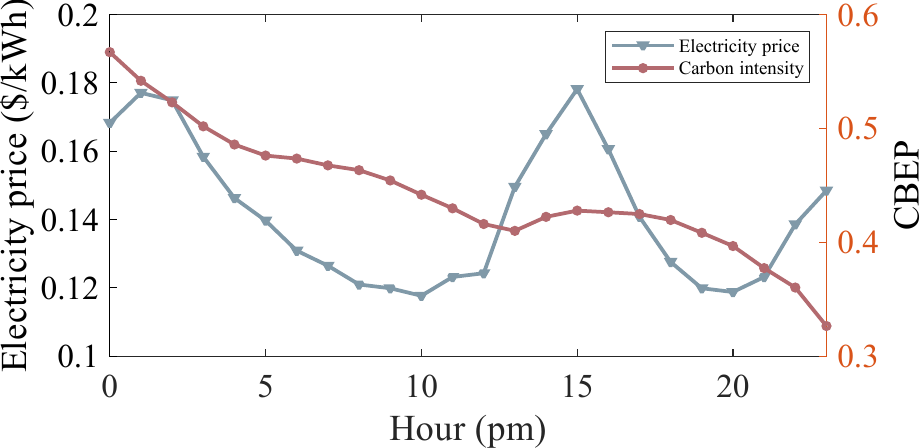}
\caption{Day-ahead carbon-based energy proportion and electricity price}
\label{fig:carbon}
\end{figure}

The hourly PV generation data are generated based on historical weather data as described by \cite{staffell2016using}. Electricity prices and hourly carbon-based energy proportion (CBEP) data, which are obtained from the Electricity Map \cite{electricitymaps}, are presented in Figure \ref{fig:carbon}. The distribution of inflexible workloads is normalized according to the historical workloads provided by \cite{luo2013temporal}. Additionally, three types of flexible jobs are assumed in the experiments based on \cite{li2016toward}.

In our numerical studies, we develop neural network models to predict day-ahead photovoltaic (PV) power generation. We use data from January 1, 2010, to December 31, 2019, comprising a total of 3,285 samples. The dataset is divided into two segments: 50\% for training and 50\% for testing. Specifically, we utilize two widely adopted prediction models: the multi-layer perceptron (MLP) and the one-dimensional convolutional neural network (1D-CNN). These models utilize input features comprising photovoltaic (PV) generation from the previous 24 hours, predicted day-ahead surface irradiance data (weighted by land area), and temporal features such as the day of the year. Based on these inputs, the models generate forecasts of PV generation for the subsequent day. Detailed model parameters are presented in Table \ref{tab:hyper}.


\begin{table}[t]
\centering
\caption{Relevant parameters for the data center}
\label{tab:g1}
\centering

\begin{tabular}{p{2.4cm}<{\centering}p{1.3cm}<{\centering}p{2.4cm}<{\centering}p{1.3cm}<{\centering}}
\toprule
Parameters& Values & Parameters &Values\\
\midrule
$P^{\text{idle}}$ (kW)& 0.1 & $P^{\text{peak}}$ (kW)& 0.2 \\
PUE & 1.4& $\mu_{d}$ $A_{max}$  & 8000 \\
  $P^{\text{ess}}_{\text{max}}$ (kW)& 80 &$Q_{ess,max}$ (kWh) & 500 \\
  $L^{\text{rate}}$& 3 &$h_c$ (h)& [2,5,7] \\
   $C^{\text{DT}}$(s)& 0.5 &$P_{max}^{\text{grid}}$ & 1000kW \\
\bottomrule
\end{tabular}
\end{table}

\begin{table}[t]
\centering
\caption{The hyper-parameters for quantile prediction}
\label{tab:hyper}
\centering

\begin{tabular}{p{2.4cm}<{\centering}p{1.3cm}<{\centering}p{2.4cm}<{\centering}p{1.3cm}<{\centering}}
\toprule
Hyper-parameters& Values & Hyper-Parameters &Values\\
\midrule
Optimizer & AdamW & Active function &ReLU \\
No. of epochs & 300& Hidden layers & [256,256] \\
No. of layers & 3 &Learning rate & 1e-3\\
Quantile-low& 0.1 & Quantile-high& 0.9  \\
\bottomrule
\end{tabular}
\end{table}

\subsection{Competing Methods}
The methods proposed in Sections III and IV are referred to as AMV-CQR and IMV-CQR. To evaluate the performance of these approaches, they are compared against other commonly used methods for decision-making under uncertainty:

1) \textit{CVaR-based Chance-Constrained Optimization (CVaR-CC)}:
This method employs Conditional Value-at-Risk (CVaR) as an approximation, which is widely used in data-driven scenarios due to its convexity and computational tractability \cite{zhai2022data}.

2) \textit{Robust Optimization with Adjusted Uncertainty Set (RO-A)}:
This approach utilizes a traditional box-shaped uncertainty set, with its size adjusted to correspond to the sample coverage based on the violation rate \cite{ben2009robust}.

3) \textit{Point-Prediction-Based Conformal Prediction (AMV-Point and IMV-Point)}: This approach adheres to the proposed CP-based framework; however, in this method, the prediction and quantile regression are conducted separately. Therefore, the conformal prediction intervals for the upper and lower bounds are identical \cite{lei2018distribution}.
\begin{table}[t]
    \centering
    \begin{tabular}{p{0.9cm}<{\centering}p{0.9cm}<{\centering}p{0.9cm}<{\centering}p{0.9cm}<{\centering}p{0.9cm}<{\centering}p{0.9cm}<{\centering}p{0.9cm}<{\centering}}
        \toprule
        $1-\alpha$ & RO-A  & CVaR-CC & AMV-Point  & IMV-Point &  \textbf{AMV-CQR}  &  IMV-CQR \\
        \midrule
        0.8 & 3449.49 & 3424.95 & 3224.68 & 3288.16 & \textbf{3213.32} & 3276.53 \\
        0.85 & 3476.73 & 3430.78 & 3252.29 & 3306.28 & \textbf{3237.16} & 3296.54 \\
        0.9 & 3505.23 & 3437.65 & 3288.94 & 3330.01 & \textbf{3272.28} & 3321.80 \\
        0.95 & 3541.06 & 3447.07 & 3341.19 & 3369.76 & \textbf{3326.67} & 3366.89 \\
        \bottomrule
    \end{tabular}
    \caption{Total cost of the data center using MLP prediction model with different methods and coverage levels (\$)}
    \label{tab:cost}
\end{table}

\begin{table}[t]
    \centering
    \begin{tabular}{p{0.9cm}<{\centering}p{0.9cm}<{\centering}p{0.9cm}<{\centering}p{0.9cm}<{\centering}p{0.9cm}<{\centering}p{0.9cm}<{\centering}p{0.9cm}<{\centering}}
        \toprule
        $1-\alpha$ & RO-A  & CVaR-CC & AMV-Point  & IMV-Point &  \textbf{AMV-CQR}  &  IMV-CQR \\
        \midrule
        0.8 & 8490.36 & 8455.52 & 7907.31 & 8066.81 & \textbf{7881.60} & 8038.85 \\
        0.85 & 8556.48 & 8470.09 & 7974.70 & 8111.44 & \textbf{7939.69} & 8088.43 \\
        0.9 & 8625.42 & 8487.76 & 8064.33 & 8169.72 & \textbf{8025.41} & 8150.57 \\
        0.95 & 8711.90 & 8511.50 & 8192.42 & 8268.60 & \textbf{8158.45} & 8261.78 \\
        \bottomrule
    \end{tabular}
    \caption{Carbon-based energy consumption using MLP prediction model with different methods and coverage levels (kWh)}
    \label{tab:carbon_emissions}
\end{table}
\subsection{Experiments}
1)\textit{ Compared with benchmarks.}
In this subsection, we perform a comparative analysis of the proposed approach against several methods commonly used in the existing literature. Tables \ref{tab:cost} and \ref{tab:carbon_emissions} present the total cost and carbon-based energy consumption, respectively, for different approaches across various coverage levels ($1-\alpha$). The conformal prediction model utilized in this analysis is the MLP Prediction Model.
The results indicate that the AMV-CQR and IMV-CQR methods achieve significant cost reductions compared to other approaches, such as RO and CVaR-CC, with maximum reductions of 6.65\% and 5.23\%, respectively, when $1-\alpha$ is 0.9. Similarly, reductions in carbon-based energy consumption are observed, with decreases of 6.96\% and 5.51\%, respectively. The conservativeness of the RO and CVaR-CC methods arises from their reliance on restrictive assumptions about the shape of the uncertainty set and their inability to incorporate covariate information (e.g., features of the prediction model such as weather, weekday, and month information). This limitation highlights the advantage of the proposed methods in leveraging additional contextual information to achieve better performance.

The results indicate that the AMV-CQR and IMV-CQR methods outperform the point-based conformal prediction methods, AMV-Point and IMV-Point. This can be attributed to the fact that, in point-based conformal prediction, prediction and quantile regression are conducted separately, resulting in identical conformal prediction intervals for the upper and lower bounds.  Furthermore, the AMV-Point and IMV-Point methods achieve superior results compared to CVaR-CC and RO, with cost improvements of 6.17\% and 5.00\%, respectively. Additionally, the findings demonstrate that uncertainty set construction methods that incorporate covariate information outperform those that do not leverage such additional information.

Table \ref{tab:vio} displays the violation rates for different methods. Average violation rates are reported for average marginal CP methods, while maximum violation rates across constraints are reported for individual CP methods and other benchmarks.
The results show that all methods successfully maintain statistical feasibility across all tested confidence levels. Examining the specific values reveals trade-offs in conservatism. CVaR-CC exhibits very low violation rates (e.g., 0.01 at $1-\alpha$ = 0.95 and $1-\alpha$ = 0.9), indicating a highly conservative approach that ensures feasibility but likely compromises solution quality. While RO-A precisely meets its target violation rate, its covariate-independent uncertainty set is static and does not dynamically adapt to changing real-time conditions affecting prediction uncertainty, leading to higher costs. Our proposed methods, AMV-CQR and IMV-CQR, effectively satisfy the strict statistical feasibility guarantees by keeping violation rates below $\alpha$ (e.g., 0.03 and 0.04 at $1-\alpha$ = 0.95). By leveraging covariate information, they provide less conservative uncertainty sets, enabling the cost and carbon-based energy reduction while maintaining required feasibility.

We also evaluated different quantile regression models for conformal prediction. Tables \ref{tab:cost_1dcnn} and \ref{tab:carbon_1dcnn} present total cost and carbon-based energy consumption, respectively, when using a 1DCNN model for prediction across various coverage levels $1-\alpha$. Comparative analysis with an MLP-based conformal prediction model revealed no significant overall performance difference; each model excelled in certain cases. This suggests that both 1DCNN and MLP sufficiently leverage covariate information to characterize prediction errors in this scenario.


\begin{table}[t]
    \centering
    \begin{tabular}{p{0.9cm}<{\centering}p{0.9cm}<{\centering}p{0.9cm}<{\centering}p{0.9cm}<{\centering}p{0.9cm}<{\centering}p{0.9cm}<{\centering}p{0.9cm}<{\centering}}
        \toprule
        $1-\alpha$ & RO-A  & CVaR-CC & AMV-Point  & IMV-Point &  \textbf{AMV-CQR}  &  IMV-CQR \\
        \midrule
        0.80  & 0.20  & 0.03  & 0.09  & 0.13  & 0.10  & 0.13  \\
        0.85 & 0.15 & 0.02 & 0.07 & 0.10  & 0.08  & 0.10  \\
        0.90  & 0.10  & 0.01 & 0.05  & 0.07  & 0.06& 0.07  \\
        0.95 & 0.05  & 0.01  & 0.03  & 0.04  & 0.03  & 0.04 \\
        \bottomrule
    \end{tabular}
    \caption{Probability of constraint violation in the chance-constraints under different methods
}
    \label{tab:vio}
\end{table}

\begin{table}[t]
    \centering
    \begin{tabular}{p{0.9cm}<{\centering}p{0.9cm}<{\centering}p{0.9cm}<{\centering}p{0.9cm}<{\centering}p{0.9cm}<{\centering}p{0.9cm}<{\centering}p{0.9cm}<{\centering}}
        \toprule
        $1-\alpha$ & RO-A  & CVaR-CC & AMV-Point  & IMV-Point &  \textbf{AMV-CQR}  &  IMV-CQR \\
        \midrule
        0.8 & 3449.49 & 3424.95 & 3224.68 & 3286.64 & \textbf{3218.48} & 3280.06 \\
        0.85 & 3476.73 & 3430.78 & 3253.22 & 3305.96 & \textbf{3241.74} & 3299.66 \\
        0.9 & 3505.23 & 3437.65 & 3287.71 & 3330.44 & \textbf{3274.64} & 3327.20 \\
        0.95 & 3541.06 & 3447.07 & 3339.55 & 3366.16 & \textbf{3328.24} & 3365.87 \\
        \bottomrule
    \end{tabular}
    \caption{Total cost of the data center 
 using 1DCNN prediction model with different methods and coverage levels (\$)}
    \label{tab:cost_1dcnn}
\end{table}

\begin{table}[t]
    \centering
    \begin{tabular}{p{0.9cm}<{\centering}p{0.9cm}<{\centering}p{0.9cm}<{\centering}p{0.9cm}<{\centering}p{0.9cm}<{\centering}p{0.9cm}<{\centering}p{0.9cm}<{\centering}}
        \toprule
        $1-\alpha$ & RO-A  & CVaR-CC & AMV-Point  & IMV-Point &  \textbf{AMV-CQR}  &  IMV-CQR \\
        \midrule
        0.8 & 8490.36 & 8455.52 & 7907.87 & 8063.81 & \textbf{7894.25} & 8047.22 \\
        0.85 & 8556.48 & 8470.09 & 7977.50 & 8111.18 & \textbf{7950.95} & 8095.45 \\
        0.9 & 8625.42 & 8487.76 & 8061.80 & 8171.29 & \textbf{8031.18} & 8163.57 \\
        0.95 & 8711.90 & 8511.50 & 8188.86 & 8260.56 & \textbf{8162.26} & 8259.00 \\
        \bottomrule
    \end{tabular}
    \caption{Carbon-based energy consumption for 1DCNN prediction model with different methods and coverage levels (kWh)}
    \label{tab:carbon_1dcnn}
\end{table}

2)\textit{ Impacts of T-EAC price.}
In this subsection, the performance of carbon-aware scheduling algorithms is assessed across different T-EAC pricing schemes. Table \ref{tab:cost_dif_price} and Table \ref{tab:carbon_emissions_dif_price} also present the relationship between total cost and carbon-based energy consumption under different T-EAC prices, respectively. The results reveal a clear correlation between increased T-EAC prices and decreased carbon-based energy consumption, indicating the effectiveness of implementing appropriate carbon pricing mechanisms in curbing environmental pollution.   It can be clearly observed
that compared with performance benchmarks, the proposed AMV-CRQ and IMV-CRQ methods can lead to about 6\% cost and carbon-based energy consumption, respectively, under any T-EAC price.

\begin{table}[t]
    \centering
    \begin{tabular}{p{0.9cm}<{\centering}p{0.9cm}<{\centering}p{0.9cm}<{\centering}p{0.9cm}<{\centering}p{0.9cm}<{\centering}p{0.9cm}<{\centering}p{0.9cm}<{\centering}}
        \toprule
        $\lambda_{c}$ &RO-A  & CVaR-CC & AMV-Point  & IMV-Point &  \textbf{AMV-CQR}  &  IMV-CQR \\
        \midrule
        0.00 & 2638.12 & 2584.21 & 2477.95 & 2508.49 & \textbf{2465.18}  & 2502.19\\
        0.05 & 3072.36 & 3011.66 & 2884.12 & 2919.93 & \textbf{2869.41}  & 2912.68\\
        0.10 & 3505.23 & 3437.65 & 3288.94 & 3330.01 & \textbf{3272.28}  & 3321.80\\
        0.20 & 4365.37 & 4284.27 & 4092.97 & 4144.58 & \textbf{4072.41}  & 4134.46\\
        \bottomrule
    \end{tabular}
    \caption{Total cost of the datacenter with different methods and T-EAC prices (\$)}
    \label{tab:cost_dif_price}
\end{table}

\begin{table}[t]
    \centering
       \begin{tabular}{p{0.9cm}<{\centering}p{0.9cm}<{\centering}p{0.9cm}<{\centering}p{0.9cm}<{\centering}p{0.9cm}<{\centering}p{0.9cm}<{\centering}p{0.9cm}<{\centering}}
        \toprule
        $\lambda_{c}$ &RO-A  & CVaR-CC & AMV-Point  & IMV-Point &  \textbf{AMV-CQR}  &  IMV-CQR  \\
        \midrule
        0.00 & 8687.14 & 8555.26 & 8125.69 & 8231.08 & \textbf{8086.77}  & 8211.93\\
        0.05 & 8681.87 & 8544.22 & 8120.78 & 8226.16 & \textbf{8081.85}  & 8207.01\\
        0.10 & 8625.42 & 8487.76 & 8064.33 & 8169.72 & \textbf{8025.41}  & 8150.57\\
        0.20 & 8579.65 & 8446.83 & 8018.61 & 8123.99 & \textbf{7979.68}  & 8104.85\\
        \bottomrule
    \end{tabular}
    \caption{Carbon-based energy consumption with different methods and T-EAC prices (kWh)}
    \label{tab:carbon_emissions_dif_price}
\end{table}

\begin{figure}[t]
\centering
\includegraphics[width=1\columnwidth]{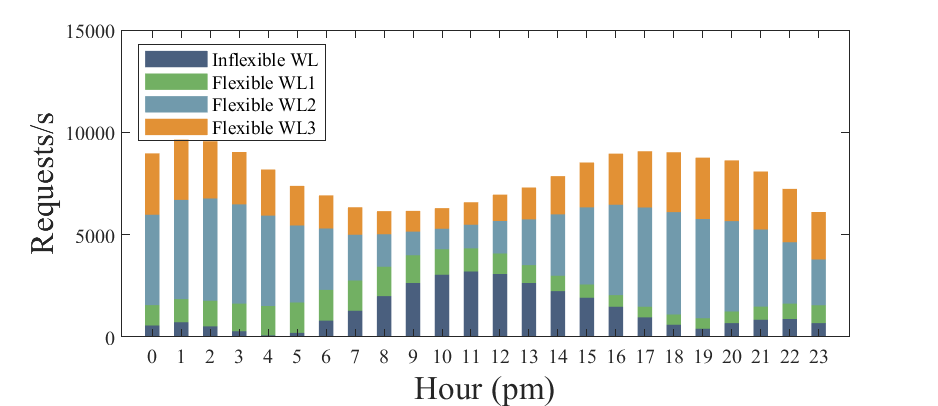}
\caption{Pattern of different classes of workloads}
\label{fig:WL}
\end{figure}
\begin{figure}[t]
\centering
\includegraphics[width=1\columnwidth]{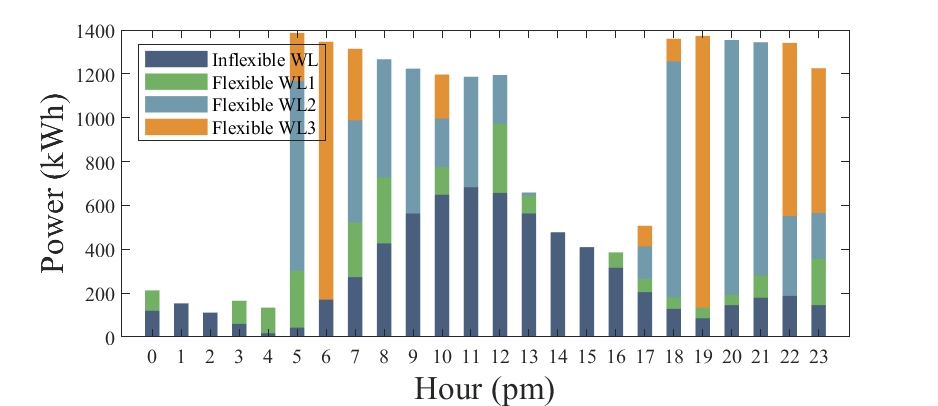}
\caption{Power consumption of different classes of workloads under carbon-aware scenario ($\lambda_{c}=\$0.1/$kWh)}
\label{fig:power_WL}
\end{figure}

3) \textit{Energy consumption of different flexible workloads}

The simulation was configured with a T-EAC price of \$0.1/kWh and a risk level of 0.1. Figure \ref{fig:power_WL} illustrates the power consumption of the workload under the carbon-aware scenario, while Figure \ref{fig:WL} presents the original workload arrival pattern. The results indicate that, within the carbon-aware scenario, flexible workloads are strategically shifted for execution to periods coinciding with lower electricity prices (hours 5–12) and periods with both low electricity prices and low grid carbon intensity (hours 18–22). Specifically, flexible workloads possessing greater time flexibility (e.g., class 3) are predominantly processed during times of low electricity prices, high renewable generation, or low carbon intensity. Conversely, flexible workloads with limited delay tolerance (e.g., class 1, with a 2-hour tolerance) are observed to be executed even during periods exhibiting high grid carbon intensity and electricity prices.

\section{Conclusion}
\label{conclusion}
Decarbonizing data centers is essential for achieving climate neutrality. In this study, a carbon-aware optimization strategy is devised to enable the efficient workload scheduling of data centers. The proposed model is designed to minimize overall energy expenses while simultaneously reducing carbon-based energy consumption toward 24/7 CEF. To address the uncertainties inherent in renewable generation predictions, a conformal prediction-based optimization framework is developed, providing statistical feasibility guarantees. The approach effectively manages uncertainties and ensures reliable and robust operation by integrating covariate information from the prediction model. The proposed framework is validated through numerical studies, demonstrating improved control over violation rates and achieving lower costs. Furthermore, the proposed approach achieves a cost reduction of over 6\% in data center workload scheduling compared to traditional methods.

\bibliographystyle{IEEEtran}
\bibliography{main}

\end{document}